\documentclass[a4paper, sort&compress]{cas-dc}

\usepackage[numbers]{natbib}
\usepackage{upgreek}
\usepackage{hyperref}
\usepackage{mhchem}
\graphicspath{{figs/}}

\def\tsc#1{\csdef{#1}{\textsc{\lowercase{#1}}\xspace}}
\tsc{WGM}
\tsc{QE}
\tsc{EP}
\tsc{PMS}
\tsc{BEC}
\tsc{DE}

\begin{document}
\begin{sloppypar}
\let\WriteBookmarks\relax
\def\floatpagepagefraction{1}
\def\textpagefraction{.001}
\shorttitle{}
\shortauthors{YX. ZHAO et~al.}

\title[mode = title]{Distinct Radiobiological Responses to BNCT in SAS Oral Squamous Cell Carcinoma and MCF-7 Breast Cancer Cells}



\author[1]{Yuxiang Zhao}[style=chinese]
\fnmark[1]

\author[1]{Zhao Sun}[style=chinese]
\author[1]{Changming Wang}[style=chinese]
\author[1]{Jianghao Lai}[style=chinese]
\author[1]{Jie Zhou}[style=chinese]
\author[1,2]{Zhencen He}[style=chinese]

\author[1]{Zhimin Hu}[style=chinese]
\cormark[1]
\ead{huzhimin@scu.edu.cn}

\affiliation[1]{
	organization={Key Laboratory of Radiation Physics and Technology of the Ministry of Education, Institute of Nuclear Science and Technology},
	addressline={Sichuan University},
	city={Chengdu},
	postcode={610064},
	country={China}
}

\affiliation[2]{
	organization={West China School of Basic Medical Sciences and Forensic Medicine},
	addressline={Sichuan University},
	city={Chengdu},
	postcode={610041},
	country={China}
}

\fntext[fn1]{This is the first author.}
	
\cortext[cor1]{Corresponding author}

\begin{abstract}
This work compared the radiobiological responses of SAS oral squamous cell carcinoma cells and MCF-7 breast cancer cells following accelerator-based boron neutron capture therapy (BNCT). Neutrons were generated by bombarding a lithium target with proton beams, followed by moderation to obtain sufficient thermal neutrons for BNCT irradiation. Boronophenylalanine (BPA) was used as the boron delivery agent. BNCT-induced biological responses were evaluated by $\gamma$-H2AX immunofluorescence staining, cell-cycle analysis, apoptosis analysis, and clonogenic survival assays. BNCT induced marked $\gamma$-H2AX foci formation in both cell lines, indicating DNA damage-associated responses after irradiation. The two cell lines further showed distinct post-irradiation outcomes. SAS cells exhibited stronger clonogenic suppression and prominent G2/M accumulation, whereas MCF-7 cells showed sustained G0/G1 accumulation and delayed apoptosis. These results suggest that BNCT sensitivity is determined by both boron accumulation and cell-line-specific biological characteristics. This work provides experimental evidence highlighting the importance of tumor-dependent cellular responses in understanding and optimizing BNCT efficacy.
\end{abstract}

\begin{highlights}
	
	\item An accelerator-based BNCT irradiation platform was established using a proton-lithium neutron source.
	
	\item BNCT induced DNA damage, apoptosis, cell-cycle changes, and reduced clonogenic survival.
	
	\item Differential BNCT sensitivity was associated with DNA damage response, cell-cycle regulation, and apoptotic susceptibility.
	
\end{highlights}


\maketitle

\section{Introduction}

Boron neutron capture therapy (BNCT) is a binary targeted radiotherapy based on the nuclear capture reaction between $^{10}$B and thermal neutrons. The $^{10}$B(n,$\alpha$)$^{7}$Li reaction produces high linear-energy-transfer (LET) $\alpha$ particles and recoiling lithium nuclei, which deposit their energy locally within boron-containing cells because of their short path length of approximately one cell diameter. This unique feature provides BNCT with the potential to selectively eliminate tumor cells while reducing damage to surrounding normal tissues \cite{sauerwein2023principles,malouff2021boron,moss2014critical,wang2022bnct}. However, the therapeutic selectivity of BNCT depends not only on neutron irradiation conditions but also on efficient boron accumulation and tumor-cell-specific biological responses \cite{barth2018boron,dymova2020boron}.

Head and neck cancers represent one of the major clinical application areas of BNCT, especially for locally advanced or recurrent tumors that are difficult to treat because of anatomical limitations \cite{aihara2006boron,kankaanranta2012boron,suzuki2014boron}. Oral squamous cell carcinoma (OSCC) is the most common malignancy of the oral cavity and is characterized by aggressive invasion, rapid proliferation, and genomic instability \cite{bray2024global,biom15050621}. Although conventional radiotherapy remains an important treatment modality for OSCC, recurrence, radioresistance, and treatment-associated limitations continue to restrict therapeutic efficacy. Previous studies have demonstrated that BNCT can inhibit OSCC cell growth and induce apoptosis, suggesting its potential for improving local tumor control \cite{kudo2020radiobiological,huang2025radiobiological}. In human OSCC models, BNCT has also been reported to induce apoptosis and regulate DNA repair-associated responses after irradiation \cite{sato2015bnct}. SAS cells, derived from human oral squamous cell carcinoma, have therefore been widely used as a model for investigating radiation-induced biological effects. In addition to their clinical relevance, SAS cells exhibit distinct molecular characteristics associated with DNA damage response regulation. Previous studies have reported TP53-associated alterations in SAS cells, which may influence checkpoint regulation and cellular responses following genotoxic stress \cite{fujita2009p53}. These characteristics make SAS cells a suitable model for evaluating how altered damage-response pathways affect BNCT-induced cellular outcomes.

Breast cancer is another common solid tumor with superficial or locally recurrent lesions for which BNCT-based therapeutic strategies have received increasing attention \cite{harbeck2019breast,seneviratne2022exploring}. From a clinical perspective, locally recurrent or chest-wall breast cancer after previous photon radiotherapy remains difficult to manage because additional irradiation is constrained by the tolerance of surrounding normal tissues. A recent clinical report of accelerator-based BNCT for recurrent breast cancer after radiotherapy further supports the potential relevance of BNCT in this setting \cite{kurosaki2024effects}. Breast cancer is biologically heterogeneous, with different molecular subtypes exhibiting distinct therapeutic responses \cite{harbeck2019breast}. MCF-7 breast cancer cells are widely used as a model for investigating radiation-induced DNA damage responses, cell-cycle regulation, and apoptosis \cite{ouyang2013genistein,wu2011ikkbeta}. Unlike SAS cells, MCF-7 cells exhibit relatively preserved p53-associated signaling and have been frequently used to study p53-dependent checkpoint activation after DNA damage \cite{casey1991growth,kastan1991cell}.Previous irradiation studies have shown that SAS and MCF-7 cells individually exhibit distinct radiation-response characteristics, suggesting that intrinsic cellular properties may influence their responses to BNCT \cite{kamida2008effect,fujita2009p53,wu2008irradiation}. However, whether these differences are maintained under accelerator-based BNCT remains unclear.

Although BNCT has demonstrated promising therapeutic potential, the biological determinants responsible for differential cellular sensitivity remain incompletely understood. BNCT generates densely ionizing radiation damage with complex DNA lesions, which may activate different cellular responses depending on DNA repair capacity, checkpoint regulation, and apoptotic susceptibility \cite{han2023calculation}. Therefore, tumor cells with different intrinsic damage-response characteristics may exhibit distinct biological outcomes following BNCT exposure. Comparing SAS and MCF-7 cells provides a biologically meaningful approach to investigate how boron accumulation and intrinsic cellular response characteristics collectively contribute to differential BNCT sensitivity.

Following BNCT irradiation, tumor cells may undergo complex DNA lesions and activate downstream biological responses, including DNA repair, cell-cycle redistribution, apoptosis, and loss of reproductive capacity \cite{maliszewska2021molecular,mavragani2017complex,punshon2024current}. Previous BNCT studies have reported the induction of DNA double-strand breaks, activation of homologous recombination repair, and changes in DNA repair-related proteins after irradiation \cite{rodriguez2018bnct}. In addition, Monte Carlo track-structure simulations have shown that DNA damage yields and DSB-based RBE in BNCT are influenced by cell size and the microscopic distribution of boron within and around cells \cite{han2023calculation}. These findings suggest that the biological consequences of BNCT are governed not only by macroscopic dose or tumor type, but also by subcellular boron distribution, cell and nuclear geometry, DNA repair capacity, checkpoint activation, and apoptotic sensitivity. Among these processes, cell-cycle redistribution is particularly important because it reflects how damaged cells delay progression, undergo repair-associated responses, or enter cell death pathways \cite{kastan2004cell,jackson2009dna}. The G1 and G2/M checkpoints restrict the progression of damaged cells into DNA replication and mitosis, respectively, and are commonly associated with p53/p21-related signaling and ATM/Chk-mediated responses after irradiation \cite{blackford2017atm,huang2020dna,shiloh2003atm,kastenhuber2017putting,matthews2022cell}.

The successful application of BNCT also depends on a suitable neutron source and an effective boron delivery strategy. Accelerator-based neutron sources have improved the accessibility, stability, and clinical feasibility of BNCT compared with traditional reactor-based systems \cite{kreiner2012accelerator,suzuki2020boron}. Boronophenylalanine (BPA) is one of the most widely used boron delivery agents because it can preferentially accumulate in tumor cells while showing clinical applicability and relatively low toxicity \cite{barth2018boron,dymova2020boron,monti2023optimizing,seneviratne2023next}. BPA is taken up by tumor cells mainly through amino acid transporter-related pathways, such as L-type amino acid transporter 1 (LAT1), which facilitate its transmembrane transport and cytoplasmic accumulation, thereby enabling intracellular $^{10}$B accumulation before neutron irradiation. Because boron uptake can differ between tumor cell types, intracellular boron accumulation should be considered when comparing BNCT sensitivity across different cellular models.

In this work, SAS oral squamous cell carcinoma cells and MCF-7 breast cancer cells were selected as two representative superficial tumor models with distinct biological characteristics.These two models represent clinically relevant tumor types with high incidence and different radiation-response backgrounds, allowing evaluation of whether BNCT responses are influenced by tissue origin and intrinsic cellular characteristics. Using BPA as the boron delivery agent and an accelerator-based neutron irradiation system, this work evaluated intracellular boron accumulation, DNA damage-associated responses, apoptosis, cell-cycle redistribution, and clonogenic survival in SAS and MCF-7 cells. This work aimed to compare the cellular responses of SAS and MCF-7 cells after BNCT exposure and to clarify how intrinsic cellular characteristics contribute to differential BNCT sensitivity.
\section{Experimental and simulation}
\subsection{Cell culture}

SAS human oral squamous cell carcinoma cells and MCF-7 human breast adenocarcinoma cells were obtained from iCell Bioscience Inc. Cells were cultured in Dulbecco's modified Eagle's medium (DMEM; Eallbio, China) supplemented with 10\% fetal bovine serum (VivaCell, Shanghai, China), 100 U/mL penicillin, and 100 $\mu$g/mL streptomycin. Cells were maintained at 37$^\circ$C in a humidified incubator containing 5\% CO$_2$. Cells in the logarithmic growth phase were used for all experiments.

\subsection{Boron compound preparation and intracellular boron analysis}

Boronophenylalanine (BPA; Sigma-Aldrich, St. Louis, MO, USA) was formulated with D-sorbitol at a ratio of 1:1.05 to improve solubility. The resulting solution was dissolved under alkaline conditions, adjusted to pH 7.2--7.4, and diluted with complete culture medium to the required working concentrations. All indicated concentrations refer to BPA concentrations in the culture medium. Cells were incubated with BPA for 24 h before irradiation or subsequent analyses.

To evaluate intracellular boron accumulation, SAS and MCF-7 cells were seeded in 6-well plates and treated with BPA at concentrations of 0, 250, 500, and 1000 $\mu$g/mL for 24 h. After incubation, cells were washed three times with Dulbecco's phosphate-buffered saline (DPBS) to remove extracellular BPA. Cells were then trypsinized, collected, counted, and subjected to microwave digestion. Boron concentrations were measured using inductively coupled plasma mass spectrometry (ICP-MS; Agilent 7700x, USA), with $^{10}$B quantified as the target isotope.

\subsection{Neutron irradiation}

Accelerator based neutrons were generated by bombarding a metallic lithium target with 4.5~MeV protons at a beam current of 10~$\mu$A using a 3~MV Tandetron accelerator at the Key Laboratory of Radiation Physics and Technology of the Ministry of Education, Institute of Nuclear Science and Technology, Sichuan University, China \cite{han2018ion}. Because of the low melting point and high chemical reactivity of lithium, a composite target structure was adopted. The target consisted of a metallic lithium layer with a diameter of 14~mm and a thickness of 450~$\mu$m deposited on a stainless steel backing with a diameter of 180~mm and a thickness of 8~mm. A water cooling system was incorporated to remove heat generated during proton irradiation.

The lithium target assembly, moderator, reflector, and cell irradiation configuration are shown in Fig.~1a. The initial neutron energy spectrum generated from the lithium target was calculated using Geant4 Monte Carlo simulations \cite{agostinelli2003geant4,allison2016recent}. Since the initially generated neutrons were mainly fast neutrons, a 100 mm thick polyethylene moderator was placed in front of the cell culture plates to reduce neutron energy and increase the thermal neutron component for BNCT irradiation. Polyethylene reflectors were arranged around the culture plates to enhance neutron utilization within the irradiation region. The simulated neutron spectra before and after moderation are shown in Fig.~1b.

After moderation and reflector optimization, the thermal neutron flux at the cell irradiation position was estimated to be $3.83 \times 10^{7}~\mathrm{n\,cm^{-2}\,s^{-1}}$. The absorbed dose components were determined by scoring the energy deposited within the irradiation volume. The total absorbed dose was defined as the sum of the boron dose ($D_B$), nitrogen dose ($D_N$), hydrogen recoil dose ($D_H$), and gamma-ray dose ($D_{\gamma}$) generated during BNCT irradiation. Based on the simulation results, the dose rates of the main BNCT dose components are summarized in Table~\ref{tab:dose_components}. The total absorbed dose was calculated as:

\begin{equation}
	D_{\mathrm{total}} = D_B + D_N + D_H + D_{\gamma},
\end{equation}

where $D_B$, $D_N$, $D_H$, and $D_{\gamma}$ represent the dose contributions from boron capture reactions, nitrogen reactions, hydrogen recoil particles, and gamma-ray irradiation, respectively.

\begin{figure*}
	\centering
	\includegraphics[width=0.8\textwidth]{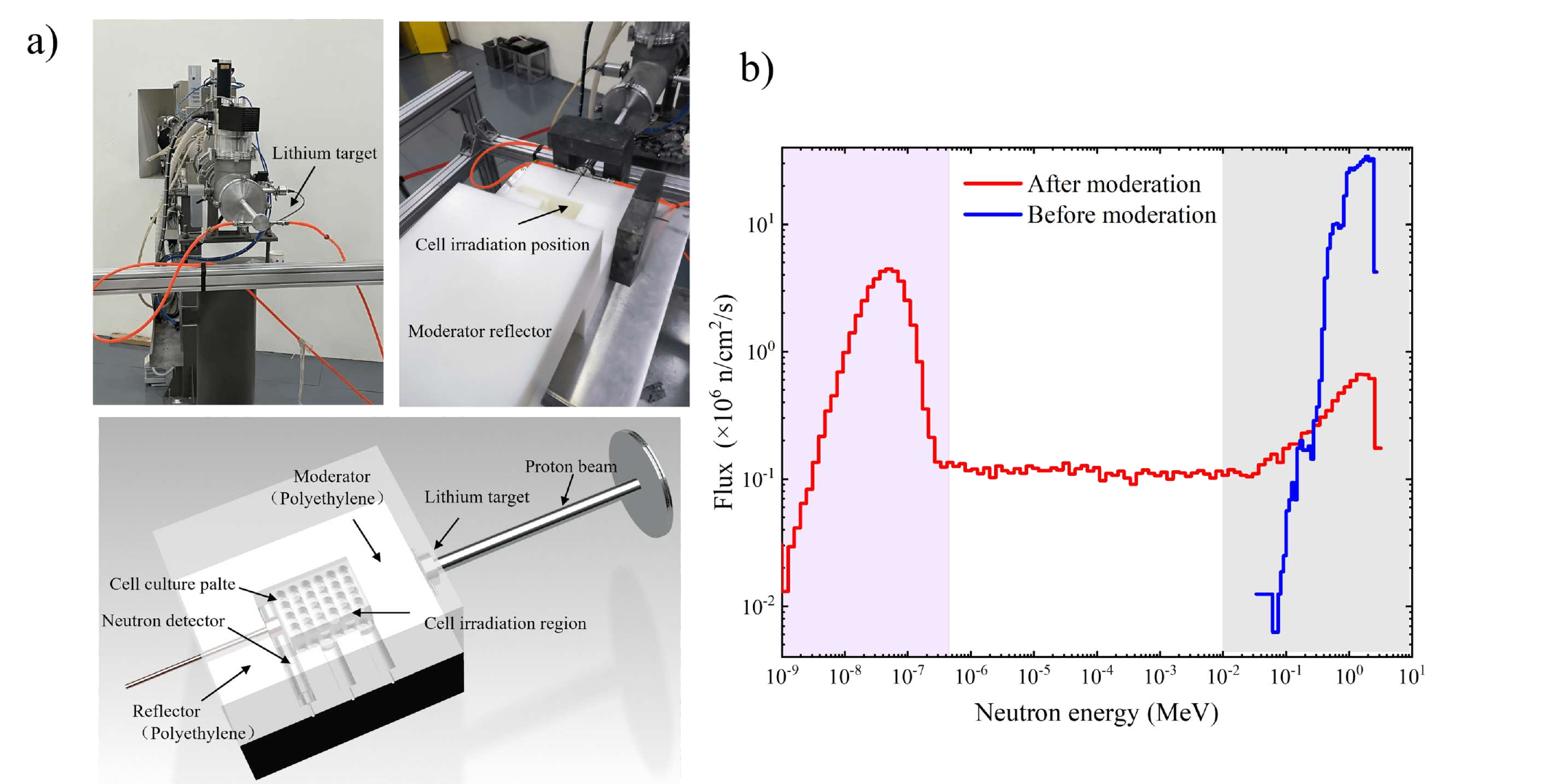}
\caption{
	Neutron irradiation setup and simulated neutron spectra for BNCT experiments.
	(a) Experimental photographs and schematic illustration of the irradiation setup, showing the lithium target assembly, moderator reflector configuration, cell culture plate, and cell irradiation region.
	(b) Geant4 simulated neutron energy spectra before and after moderation. The red curve represents the neutron spectrum after moderation, and the blue curve represents the neutron spectrum before moderation. The shaded regions indicate different neutron energy ranges.
}
	\label{sy picture}
\end{figure*}
\begin{table}[!t]
	\centering
	\caption{Simulated dose-rate components during BNCT irradiation.}
	\label{tab:dose_components}
	\begin{tabular}{lll}
		\hline
		Dose component & Symbol & Dose rate (Gy/h) \\
		\hline
		Boron dose & $D_B$ & 0.25 \\
		Nitrogen dose & $D_N$ & 0.025 \\
		Hydrogen recoil dose & $D_H$ & 0.43 \\
		Gamma-ray dose & $D_{\gamma}$ & 0.40 \\
		\hline
		Total absorbed dose rate & $D_{\mathrm{total}}$ & 1.105 \\
		\hline
	\end{tabular}
\end{table}
\subsection{CCK-8 viability assay}

Cell viability after BPA treatment was evaluated using a CCK-8 assay. SAS and MCF-7 cells were seeded in 96-well plates and treated with BPA at concentrations of 0, 250, 500, and 1000 $\mu$g/mL for 24~h. After incubation, 10~\textmu L of CCK-8 solution (Biosharp, Anhui, China) was added to each well, followed by incubation at $37^{\circ}\mathrm{C}$ for 2~h. Absorbance was measured at 450~nm using a microplate reader. Cell viability was calculated relative to the untreated control group according to the manufacturer's instructions. All experiments were performed in triplicate.

\subsection{$\gamma$-H2AX immunofluorescence staining}

To evaluate DNA damage-associated responses, $\gamma$-H2AX immunofluorescence staining was performed after BNCT irradiation. Cells were irradiated for 3~h under the BNCT condition described above, and samples were collected at 0.5, 1, 2, 4, 6, and 24~h after the end of irradiation. Cells were washed with PBS, fixed with 4\% paraformaldehyde, permeabilized with 0.5\% Triton X-100, and blocked with 1\% BSA. Cells were then incubated with anti-$\gamma$-H2AX primary antibody at $4^{\circ}$C overnight. After washing with PBS, cells were incubated with AF594-conjugated secondary antibody for 2~h in the dark. Nuclei were counterstained with DAPI, and $\gamma$-H2AX foci were visualized using fluorescence microscopy. For quantitative analysis, 50--100 nuclei were randomly selected from each group, and the number of $\gamma$-H2AX foci per nucleus was quantified using ImageJ.

\subsection{Cell cycle and apoptosis detection}

After irradiation, cells were harvested at 12, 24, 48, and 72~h for cell-cycle and apoptosis analyses. For cell-cycle analysis, cells were washed with PBS and fixed in 70\% cold ethanol overnight. Fixed cells were then washed with PBS and stained with propidium iodide (PI) solution containing RNase A. Cell-cycle distribution was analyzed by flow cytometry (BD FACS Canto II), and the percentages of cells in the G0/G1, S, and G2/M phases were quantified using FlowJo software.

For apoptosis detection, cells were harvested at the same time points and washed with cold PBS. Unlike the cell-cycle samples, apoptosis samples were not fixed with ethanol before staining. Cells were stained with Annexin V-FITC and PI according to the manufacturer's protocol (BD Biosciences). Stained cells were analyzed by flow cytometry (BD FACS Canto II), and early apoptotic, late apoptotic, and total apoptotic populations were quantified using FlowJo software.
\subsection{Colony formation assay}
After neutron irradiation for 1, 2 or 3~h, cells were seeded at different densities according to the treatment group. For MCF-7 cells, all groups were seeded at 500 cells per well. For SAS cells, untreated control and BPA-only groups were seeded at 500 cells per well; neutron-only groups at 1000 cells per well; and BNCT groups at 3000 cells per well. Cells were cultured in 6-well plates with 2~mL of complete medium per well. MCF-7 cells were incubated at $37^{\circ}$C in a 5\% CO\textsubscript{2} incubator for 8 days, and SAS cells for 14 days. After incubation, the medium was removed, and cells were washed twice with cold PBS. Fixation was performed with 70\% cold ethanol for 15~min, followed by staining with 0.5\% crystal violet in methanol for 15~min. Excess dye was removed by rinsing with distilled water, and the plates were air-dried upside down. Colonies containing at least 50 cells were counted using ImageJ or a colony counter. Plating efficiency and surviving fraction were calculated relative to untreated controls. All experiments were performed in triplicate. The formula for calculating the reproductive death rate is as follows:
\begin{equation}\label{eq:reproductive}
	R_{d} = \left( 1 - \frac{N}{N_{0}} \right) \times 100\%
\end{equation}

where $R_{d}$ is the reproductive death rate, and $N$ and $N_{0}$ are the normalized colony formation rates of the treated and untreated control groups, respectively.
\subsection{Statistical analysis}
Statistical analyses were performed using GraphPad Prism version 10.1.2 (GraphPad Software, San Diego, CA, USA). Data are presented as mean \(\pm\) standard deviation (SD). One-way analysis of variance (ANOVA) was used to evaluate differences among groups. A \(p\)-value \(< 0.05\) was considered statistically significant. Significance levels are indicated as \(*p < 0.05\), \(**p < 0.01\), \(***p < 0.001\), \(****p < 0.0001\), and n.s. for not significant.

\section{Results and analysis}

\subsection{BPA cytotoxicity and cellular boron uptake}

CCK-8 assays showed that BPA concentrations up to 500~$\mu$g/mL did not significantly affect the viability of either SAS or MCF-7 cells after 24~h incubation, with cell viability remaining above 85\% of the control level. In contrast, 1000~$\mu$g/mL BPA induced significant cytotoxicity, reducing cell viability to below 70\%. Intracellular boron concentration, as measured by ICP-MS, increased with increasing BPA concentration. After 24~h incubation with 500~$\mu$g/mL BPA, SAS cells showed a higher intracellular boron concentration than MCF-7 cells, reaching 153~$\mu$g/L compared with approximately 50~$\mu$g/L in MCF-7 cells (Fig.~2). Considering both BPA cytotoxicity and intracellular boron accumulation, 500~$\mu$g/mL BPA was selected for subsequent BNCT experiments.

\begin{figure}
	\centering
	\includegraphics[width=0.5\textwidth]{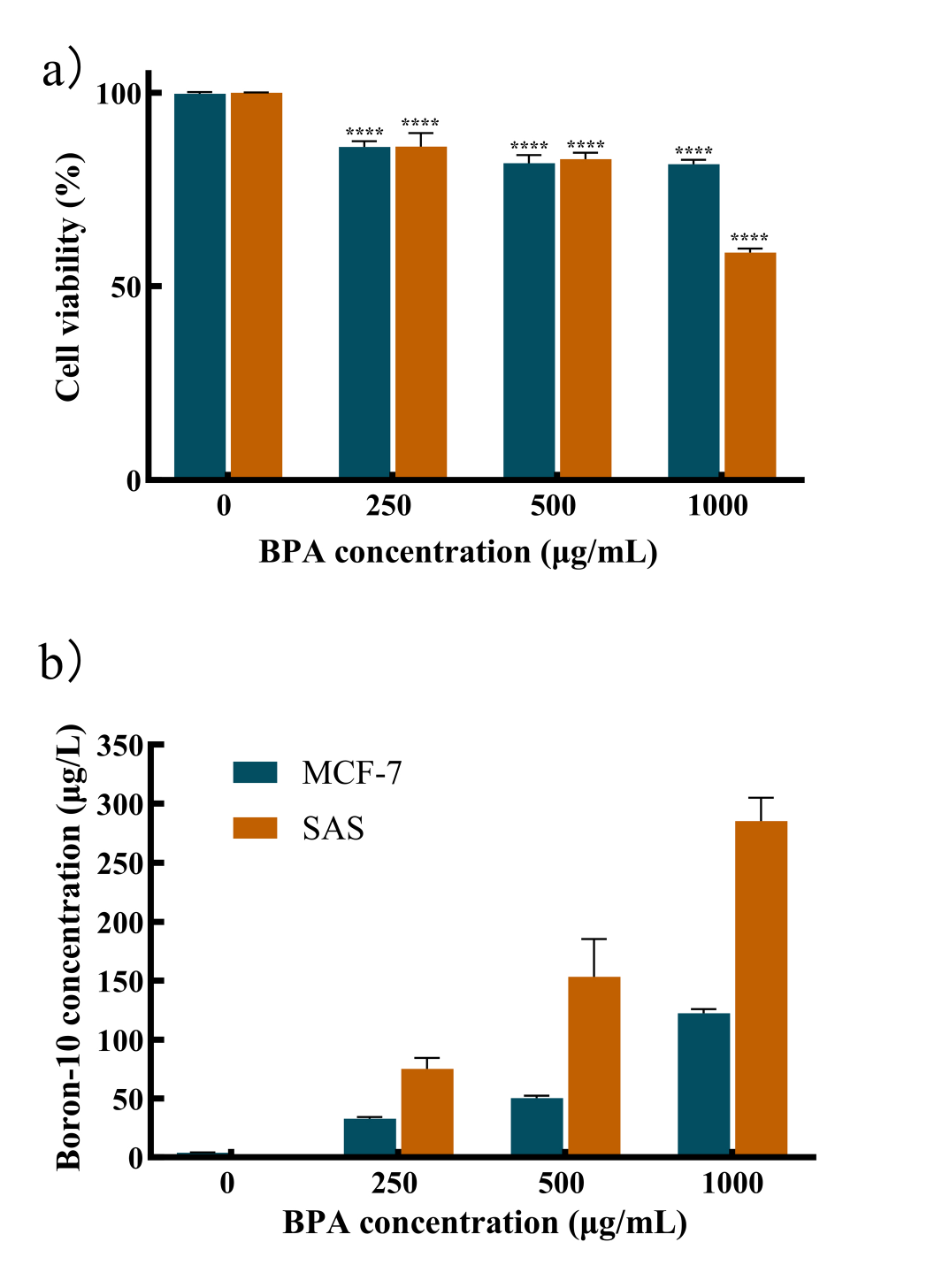}
\caption{
	BPA cytotoxicity and boron accumulation in SAS and MCF-7 cells.
	(a) Cell viability determined by CCK-8 assay after 24~h BPA treatment.
	(b) Intracellular boron concentration measured by ICP-MS after 24~h BPA treatment.
	Data are presented as mean $\pm$ SD.
}
	\label{BPA}
\end{figure}

\subsection{BNCT-induced $\gamma$-H2AX foci formation}

DNA damage-associated responses were evaluated by $\gamma$-H2AX immunofluorescence staining. BNCT induced a marked increase in $\gamma$-H2AX foci at 0.5~h after irradiation. In SAS cells, approximately 70 foci per cell were observed after BNCT, whereas approximately 18 foci per cell were detected in the neutron-only group. In MCF-7 cells, approximately 38 foci per cell were observed after BNCT, compared with approximately 14 foci per cell after neutron-only irradiation. Only minimal $\gamma$-H2AX foci formation was observed in the control and BPA-alone groups. A time-course analysis from 0.5 to 24~h showed a gradual decrease in $\gamma$-H2AX foci after BNCT irradiation. In SAS cells, the number of foci decreased from approximately 70 at 0.5~h to approximately 8 at 6~h and approached baseline levels by 24~h. In MCF-7 cells, $\gamma$-H2AX foci also decreased over time and reached low levels at 24~h (Fig.~3).
\begin{figure*}
	\centering
	\includegraphics[width=0.9\textwidth]{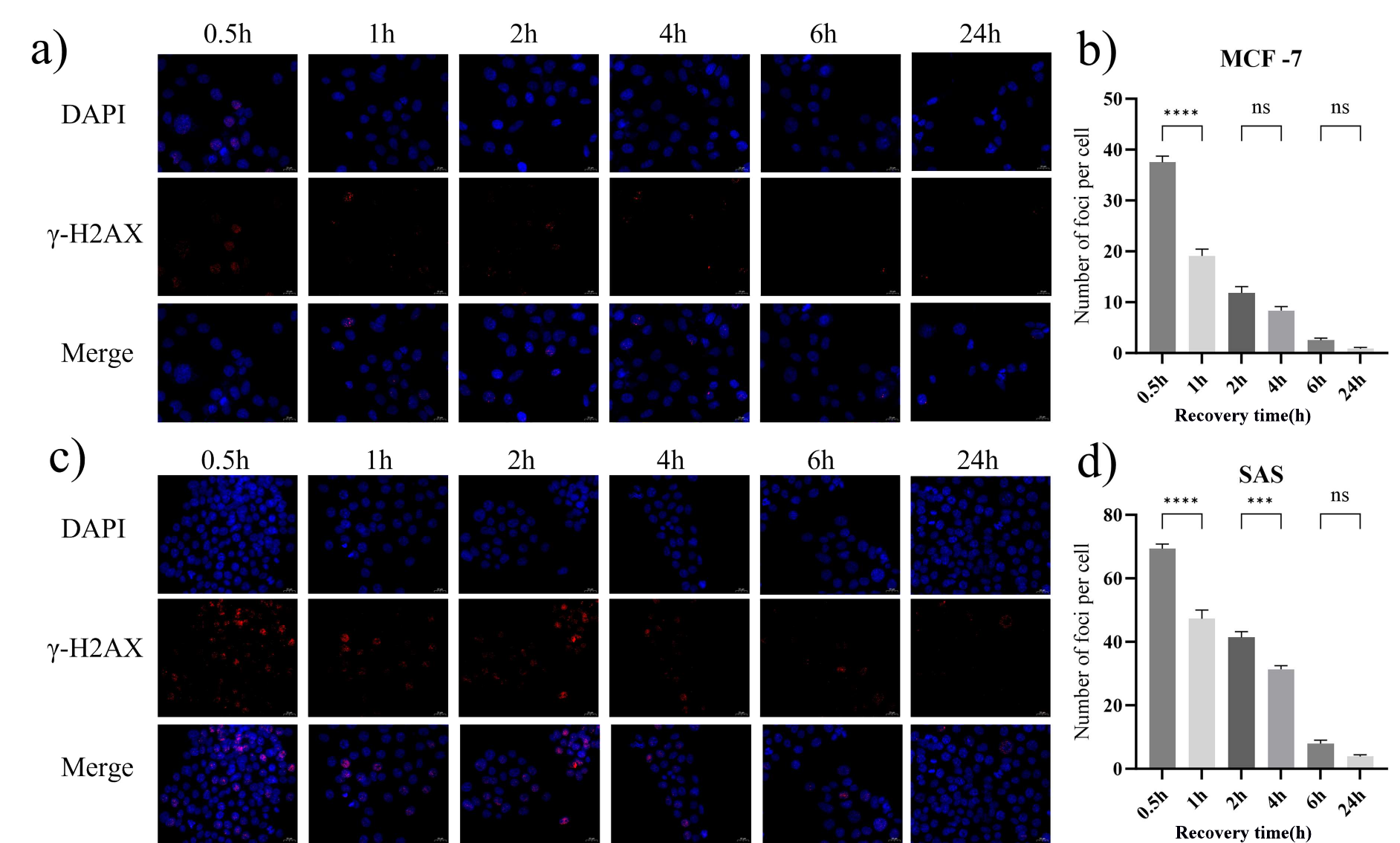}
	\caption{
		BNCT induced time-dependent changes in $\gamma$-H2AX foci formation in MCF-7 and SAS cells. 
		(a,b) Representative immunofluorescence images (a) and quantitative analysis (b) of $\gamma$-H2AX foci in MCF-7 cells from 0.5 to 24~h after BNCT irradiation. 
		(c,d) Representative immunofluorescence images (c) and quantitative analysis (d) of $\gamma$-H2AX foci in SAS cells from 0.5 to 24~h after BNCT irradiation. 
		Red: $\gamma$-H2AX; blue: DAPI-stained nuclei. 
	}
	\label{fig:gamma_h2ax}
\end{figure*}
\begin{figure*}
	\centering
	\includegraphics[width=0.8\textwidth]{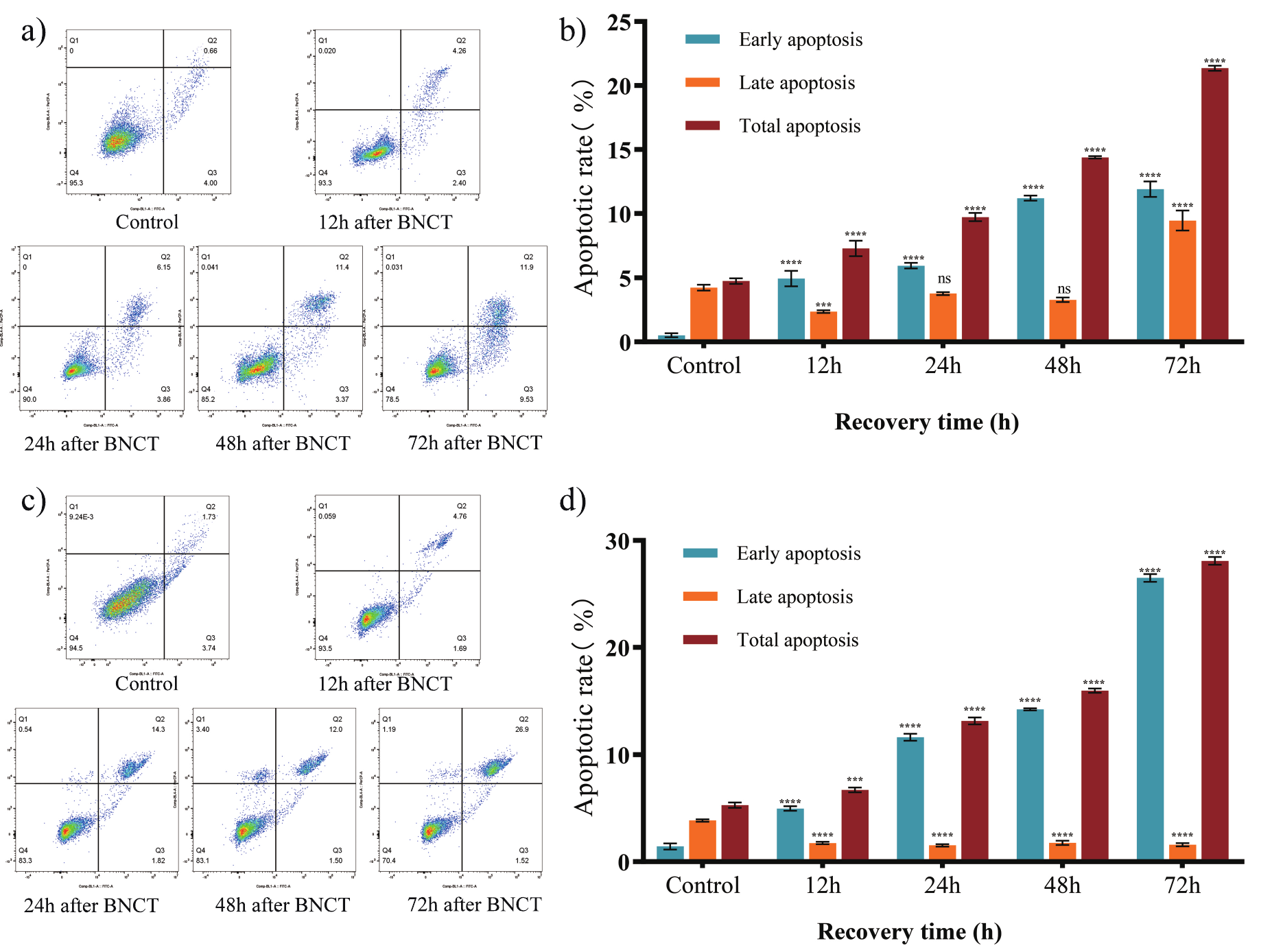}
	\caption{BNCT induces time-dependent apoptosis in SAS and MCF-7 cells. (a--d) Representative Annexin V/PI flow cytometry plots (a, c) and quantification of early, late, and total apoptotic rates (b, d) in SAS (a, b) and MCF-7 (c, d) cells from 12 to 72~h post-BNCT.}
	\label{fig:apoptosis}
\end{figure*}

\begin{figure*}
	\centering
	\includegraphics[width=0.65\textwidth]{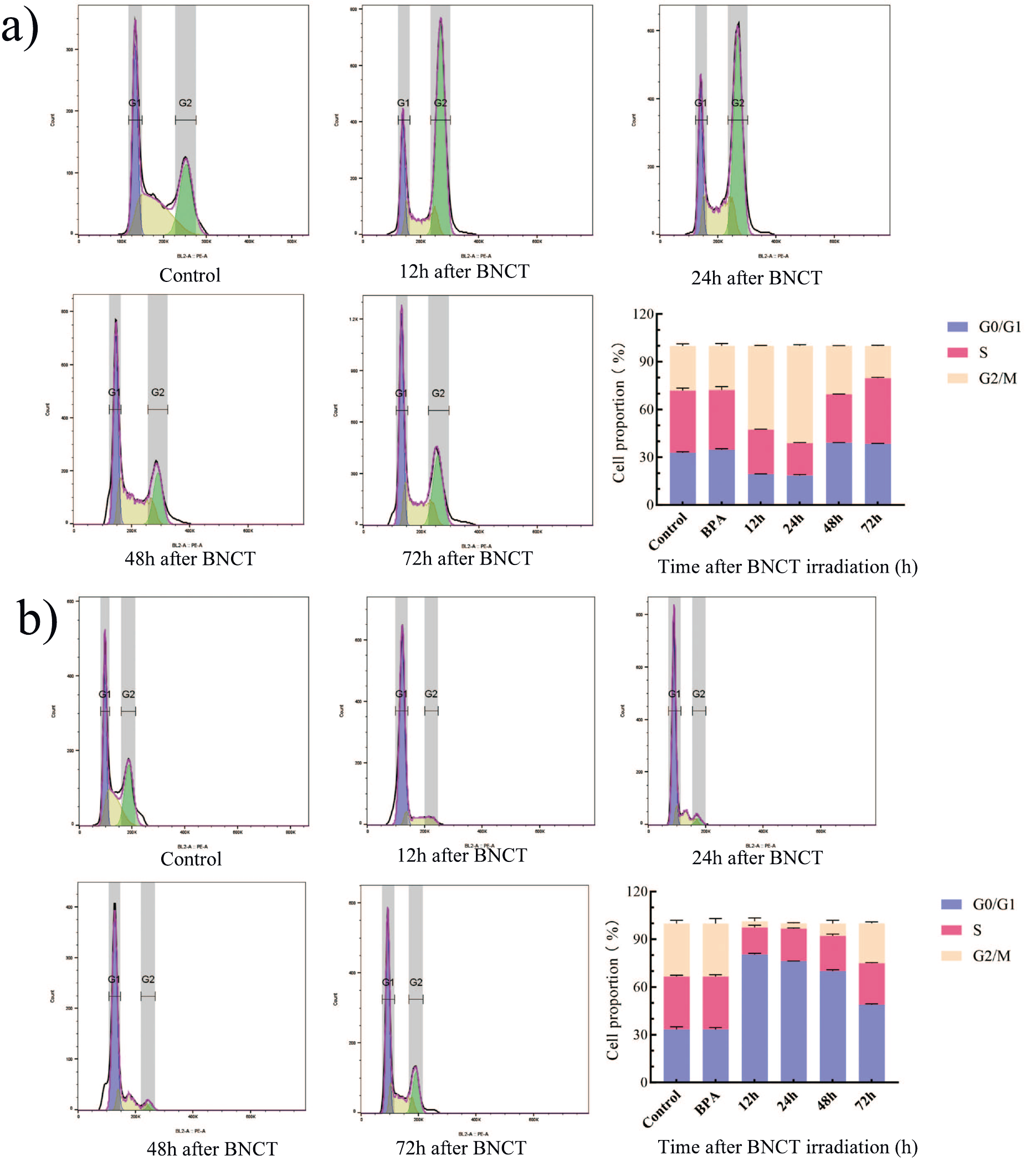}
	\caption{BNCT induces distinct cell cycle arrest patterns in SAS and MCF-7 cells. (a, b) Representative flow cytometry histograms and corresponding quantitative analysis of cell cycle phase distributions (G0/G1, S, and G2/M) in SAS (a) and MCF-7 (b) cells from 12 to 72~h post-BNCT. Control and BPA-alone groups were evaluated at 24~h. Data are presented as mean $\pm$ SD.}
	\label{fig:cell_cycle}
\end{figure*}

	\subsection{Induction of apoptosis by BNCT}
Apoptosis was quantified by flow cytometry using Annexin V/PI staining. In BNCT-treated SAS cells, early apoptosis increased rapidly from 4.9\% at 12~h to 11.2\% at 48~h and remained at 11.9\% at 72~h. Late apoptosis rose from 2.3\% at 12~h to 9.4\% at 72~h. Total apoptosis increased from 7.3\% at 12~h to 21.4\% at 72~h. In MCF-7 cells, early apoptosis gradually increased from 6.7\% at 12~h to 26.5\% at 72~h, while late apoptosis remained low (1.5\%--1.7\%). Total apoptosis increased from 6.7\% at 12~h to 28.1\% at 72~h. BPA alone did not induce apoptosis above control levels in either cell line (Fig.~{4}).

\subsection{Induction of cell cycle arrest by BNCT}
SAS and MCF-7 cells were treated with BNCT at a physical dose of 3.315 Gy and subjected to flow cytometry. In SAS cells, BNCT induced a distinct pattern: the G0/G1 phase decreased from 32.0\% in controls to 18.0\% at 12~h and 24~h, then returned to 40.0\% at 48~h and 72~h. The G2/M phase increased markedly from 28.0\% in controls to 62.0\% at 24~h, while the S phase showed a transient decline at 12~h (38.0\%) followed by a gradual increase. Conversely, in MCF-7 cells, the proportion of cells in the G0/G1 phase increased dramatically from 33.3\% in controls to 80.3\% at 12~h after BNCT, remained elevated at 24~h (76.2\%) and 48~h (69.9\%), and then declined to 50.0\% at 72~h. The S phase fraction decreased from 33.3\% to 17.0\% at 12~h and remained reduced thereafter. G2/M phase cells decreased from 33.3\% to 2.5--4.5\% between 12 and 72~h (Fig.~{5}). BPA alone had no significant effect on cell cycle distribution in either cell line.

	\subsection{Suppression of colony formation by BNCT}
Clonogenic survival was determined after 1, 2 or 3~h irradiation. 
In both cell lines, survival fractions decreased with increasing irradiation time, but SAS cells were suppressed much more strongly than MCF-7 cells. After 1~h of BNCT, the survival fractions of SAS and MCF-7 cells were 15.2\% and 47.3\%, respectively. Extending the irradiation to 3~h further reduced survival to 10.1\% in SAS cells and 45.1\% in MCF-7 cells (Fig.~{6}.). Neutron-only irradiation reduced survival to a much lesser extent. The BNCT-specific increment in cell killing was much larger in SAS cells (~70\% additional reproductive death at 1~h) compared to MCF-7 cells (~29\% additional death at 1~h). Notably, the survival curves for BNCT plateaued after 1--2~h.
\begin{figure*}
	\centering
	\includegraphics[width=0.6\textwidth]{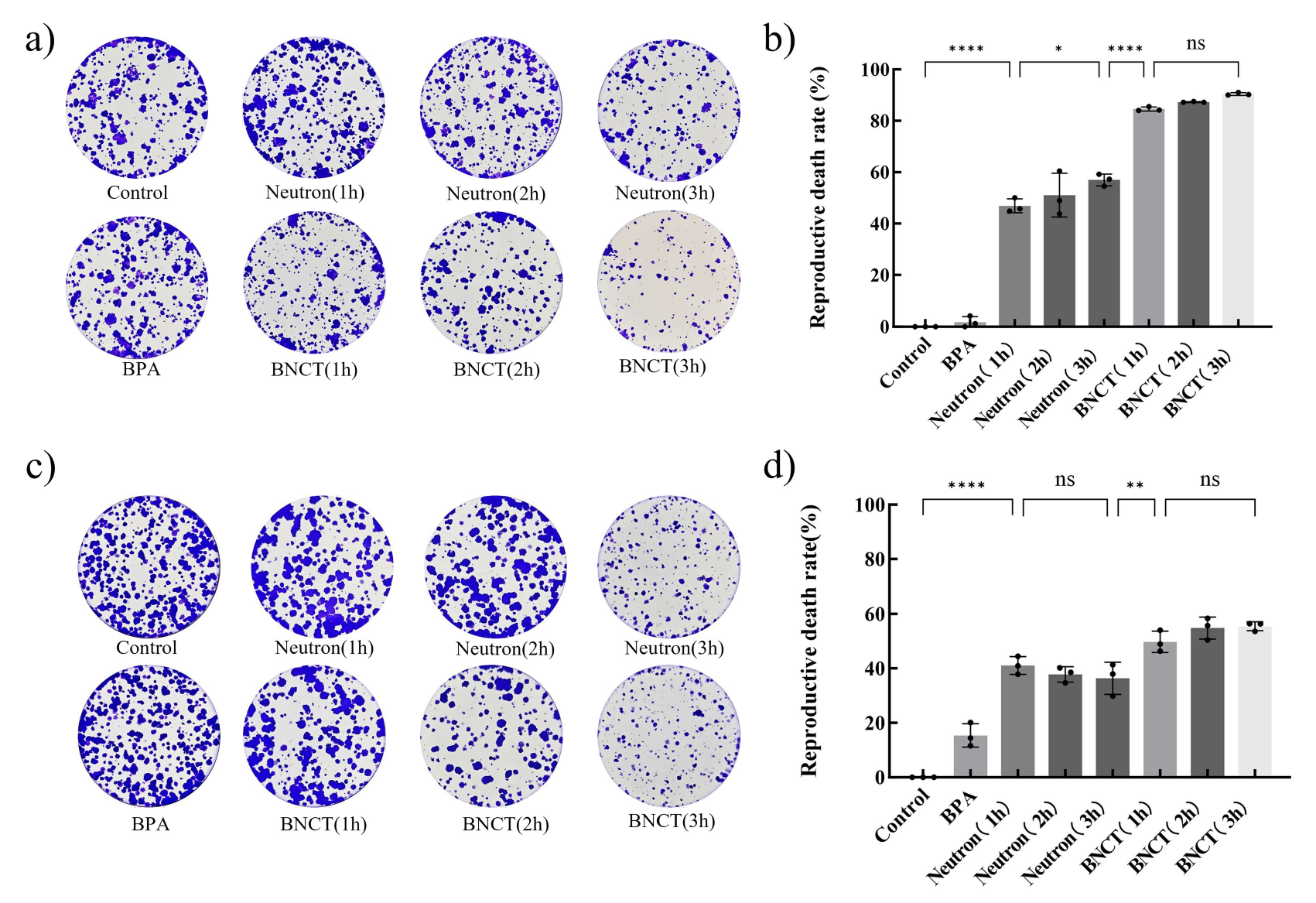}
	\caption{BNCT strongly suppresses colony formation and induces reproductive death in SAS and MCF-7 cells. (a, c) Representative images of clonogenic survival assays for SAS (a) and MCF-7 (c) cells following 1, 2, and 3~h of neutron-only or BNCT irradiation. (b, d) Corresponding quantification of reproductive death rates in SAS (b) and MCF-7 (d) cells. Control and BPA-alone groups were included for comparison. Data are presented as mean $\pm$ SD.}
	\label{fig:clone}
\end{figure*}

\section{Discussion}

This work demonstrated that SAS oral squamous cell carcinoma cells and MCF-7 breast cancer cells exhibited distinct responses to BNCT using BPA as the boron delivery agent. SAS cells showed higher intracellular boron accumulation, stronger early $\gamma$-H2AX foci formation, more pronounced clonogenic suppression, and a G2/M-dominant cell-cycle response, whereas MCF-7 cells showed lower boron accumulation, sustained G0/G1 accumulation, delayed apoptosis, and relatively higher clonogenic survival. These findings indicate that the biological effects of BNCT are influenced not only by boron uptake but also by the intrinsic biological responses of different tumor cells after irradiation. The stronger early $\gamma$-H2AX response in SAS cells may be partly related to higher intracellular boron accumulation and greater microscopic dose deposition. Although $\gamma$-H2AX foci decreased over time in both cell lines, this attenuation should not be interpreted as complete DNA repair, because residual damage or downstream cellular consequences may persist after the decline of $\gamma$-H2AX signals \cite{roos2013dna,todorovic2019mechanisms}.

The different cell-cycle and apoptosis patterns further suggest distinct downstream responses following BNCT-induced DNA damage-associated signaling. In SAS cells, BNCT induced a G2/M-dominant response at early time points, indicating that damaged cells were mainly delayed before mitotic entry. The subsequent decrease in G2/M accumulation at 48--72~h temporally coincided with a marked increase in apoptosis. Notably, late apoptosis became more evident after 48~h, suggesting that a fraction of damaged G2/M-arrested cells may have progressed toward irreversible apoptotic cell death rather than fully recovering normal cell-cycle progression. This pattern is consistent with G2/M checkpoint-associated responses after irradiation, including ATM/Chk-related signaling and Cyclin B1/CDK1 regulation \cite{blackford2017atm,huang2020dna,matthews2022cell}.

In contrast, MCF-7 cells showed a G0/G1-dominant response after BNCT irradiation, suggesting preferential delay before DNA replication. The partial decline of G0/G1 accumulation at later time points was accompanied by increased Annexin V positivity, indicating delayed apoptotic responses after sustained cell-cycle arrest. However, late apoptosis in MCF-7 cells remained relatively low and did not show a marked time-dependent increase, suggesting that apoptotic progression was mainly limited to early apoptotic changes within the observation period. The G0/G1-dominant pattern is commonly associated with p53-p21-related G1/S checkpoint control after DNA damage \cite{kastenhuber2017putting,engeland2022cell,hafner2019multiple}. Together, these results suggest that the apparent recovery of cell-cycle distribution in both cell lines may partly reflect apoptosis-associated elimination of damaged, cell-cycle-arrested cells rather than complete restoration of normal cell-cycle progression. This interpretation is consistent with the clonogenic survival results: SAS cells showed increased late apoptosis and severe clonogenic suppression, whereas MCF-7 cells showed delayed but mainly early apoptotic responses and relatively higher clonogenic survival. Therefore, apoptosis and reproductive death should be interpreted as related but not equivalent endpoints, and the overall BNCT response should be evaluated as the integrated outcome of DNA damage-associated signaling, cell-cycle redistribution, apoptotic progression, and loss of clonogenic capacity.

The present findings are partly consistent with previous BNCT studies in squamous cell carcinoma models. Kamida et al. reported that BNCT with BPA as the boron delivery agent inhibited colony formation in SAS cells and induced cell-cycle checkpoint responses together with apoptosis \cite{kamida2008effect}. Fujita et al. further showed that p53 status influenced BNCT sensitivity in oral squamous cell carcinoma cells, with both G1- and G2-related checkpoint responses contributing to cellular outcomes \cite{fujita2009p53}. A recent AB-BNCT study in SAS cells also reported G2/M arrest, apoptosis, and strong clonogenic suppression after neutron irradiation with BPA as the boron delivery agent \cite{huang2025radiobiological}. Together with the present results, these findings suggest that BNCT sensitivity is influenced by the combined effects of boron uptake, DNA damage-associated signaling, cell-cycle redistribution, apoptotic progression, and reproductive survival. Future work should further validate the checkpoint-related molecular pathways underlying cell-line-dependent BNCT responses.

\section{Conclusion}

In this work, the responses of SAS oral squamous cell carcinoma cells and MCF-7 breast cancer cells to BNCT were compared using BPA as the boron delivery agent and an accelerator-based neutron irradiation system. Marked $\gamma$-H2AX foci formation was observed in both cell lines after irradiation, indicating DNA damage-associated responses. However, SAS cells showed stronger early DNA damage signals, prominent G2/M accumulation, and severe clonogenic suppression, whereas MCF-7 cells exhibited sustained G0/G1 accumulation and delayed Annexin V positivity.

These results suggest that BNCT sensitivity is determined by the combined effects of boron uptake, DNA damage induction, cell-cycle redistribution, apoptosis, and reproductive survival. Because checkpoint proteins were not directly examined, the molecular mechanisms underlying these cell-line-dependent responses require further validation. This work highlights the importance of considering tumor cell-specific biological characteristics when evaluating and optimizing BNCT efficacy.

\section*{Acknowledgements}
This work was supported by the Sichuan Science and Technology Program
(Grant No. 2025NSFJQ0042).


\bibliography{cas-refs.bib}
\bibliographystyle{unsrt}
\vskip3pt
\end{sloppypar}
\end{document}